\documentclass[aps,prc,twocolumn,amsmath,amssymb,floatfix,superscriptaddress]{revtex4-2}
\usepackage{mathptmx}
\usepackage[T1]{fontenc}
\usepackage{CJK}
\usepackage{graphicx}
\usepackage{bm}
\usepackage{dcolumn}
\usepackage{xcolor}
\usepackage{epstopdf}
\usepackage{multirow} 
\usepackage{booktabs} 
\usepackage[colorlinks,allcolors=blue]{hyperref}
\usepackage{tablefootnote}   
\usepackage{orcidlink}

\renewcommand{\arraystretch}{1.5} 
\newcommand{\be}{\begin{equation}}
\newcommand{\ee}{\end{equation}}
\newcommand{\bea}{\begin{eqnarray}}
\newcommand{\eea}{\end{eqnarray}}

\begin{document}

\title{Deformed neutron halo nuclei and soft dipole excitations in the $40<A<90$ mass region}

\author{Xiao Lu\orcidlink{0000-0002-0037-7890}} 
\email[]{luxiao@itp.ac.cn}
\affiliation{Institute of Theoretical Physics, Chinese Academy of Sciences, Beijing 100190, China} 
\author{Cong Pan\orcidlink{0000-0003-3675-8238}}
\affiliation{Department of Physics, Anhui Normal University, Wuhu 241000, China}
\author{Hiroyuki Sagawa\orcidlink{0000-0002-6900-7855}} 
\email[]{sagawa@ribf.riken.jp}
\affiliation{Institute of Theoretical Physics, Chinese Academy of Sciences, Beijing 100190, China}
\affiliation{RIKEN Nishina Center for Accelerator-Based Science, Wako 351-0198, Japan} 
\affiliation{Center for Mathematics and  Physics, University of Aizu, Aizu Wakamatsu, Fukushima 965-0001, Japan} 
\author{Xiang-Xiang Sun\orcidlink{0000-0003-2809-4638}} 
\affiliation{School of Nuclear Science and Technology, University of Chinese Academy of Sciences, Beijing 100049, China}
\affiliation{Institute for Advanced Simulation (IAS-4), Forschungszentrum J\"{u}lich, D-52425 J\"{u}lich, Germany}
\author{Shan-Gui Zhou\orcidlink{0000-0003-4753-3325}} 
\email[]{sgzhou@itp.ac.cn} 
\affiliation{Institute of Theoretical Physics, Chinese Academy of Sciences, Beijing 100190, China} 
\affiliation{School of Nuclear Science and Technology, University of Chinese Academy of Sciences, Beijing 100049, China} 
\affiliation{School of Physical Sciences, University of Chinese Academy of Sciences, Beijing 100049, China}  

\date{\today}

\begin{abstract} 
We study deformed neutron halo nuclei in the mass region $40 < A < 90$ and their soft electric dipole ($E1$) excitations based on the deformed relativistic Hartree-Bogoliubov theory in continuum (DRHBc). Three candidates, $^{43}$Si, $^{69}$Ti, and $^{75}$Cr, are selected for detailed analysis. 
Unique features are identified in the decoupled densities of possible $s$- and $p$-wave deformed halo nuclei in this mass region, which are influenced by large high-$l$ configurations. It is shown that the dipole response is a highly sensitive
observable to detect the halo component of the single-particle wave function in deformed halo nucleus,  and it helps identify the configuration and the magnitude of deformation for halo nuclei in the $40 < A < 90$ mass region. Experimental confirmation of the dipole strength in the low-energy region is highly desirable to explore possible deformed halo candidates in the medium-heavy mass region. 
\end{abstract}

\maketitle
\section{Introduction}
The first experimental observation of the halo nucleus $^{11}$Li 
was reported in 1985 through high-energy interaction cross-section measurements \cite{Tanihata1985}. Following this pioneering work, numerous halo nuclei have been identified using interaction cross-section or Coulomb dissociation measurements. These include $1n$ halo nuclei such as $^{11}$Be \cite{TANIHATA1988,FUKUDA1991}, $^{19}$C \cite{19c1995,Nakamura1999}, $^{31}$Ne \cite{Nakamura2009,Nakamura2014}, and $^{37}$Mg \cite{mg37,mg371}, as well as the $2n$ halo nuclei like $^{17}$B \cite{17B02,17B21}, $^{19}$B \cite{19B20}, $^{22}$C \cite{Tanaka2010}, and $^{29}$F \cite{29F20}. Along the proton drip line, $^{8}$B \cite{PhysRevLett.69.2058} and $^{12}$N~\cite{1998N12} have been identified as $1p$ halo nuclei, while $^{17}$Ne~\cite{1998N12} has been categorized as a $2p$ halo nucleus. Notably, nuclear deformation plays a key role in the formation of the halo structure in $^{31}$Ne and $^{37}$Mg. A challenging question for both theory and experiment is whether halo structures can also emerge in heavier nuclei with $A>40$ and, if so, in which regions of the nuclear chart, especially near the neutron drip lines, these structures might be found. 

The halo is formed under the condition that the nucleus has a very small $1n$ or $2n$ separation energy, and the valence nucleon(s) occupy low angular momentum  $l\leq 1$ orbitals \cite{SAGAWA1992,RIISAGER1992,meng96,MENG06,TANIHATA2013215,Meng_2015}. In the case of spherical nuclei with $20 < N < 28$, the latter condition cannot be satisfied when adopting the conventional single-particle shell model, due to valence neutrons occupying the $1f_{7/2}$ orbit, whose large centrifugal barrier prevents halo formation. So far, the heaviest halo nucleus identified experimentally is $^{37}$Mg, in which the deformation effect plays a crucial role \cite{mg37,mg371}. It is expected that heavier nuclei with $A>40$ could be difficult to form a halo nature because more and more high-angular-momentum orbitals will be involved in the shell model configuration for spherical nuclei. Nevertheless, in medium and heavy spherical systems, giant halos have been predicted in extremely neutron-rich Ca and Zr isotopes \cite{MengJ98,Meng:2002ps,MENG06,Terasaki:2006tw}, but experimental confirmation of these exotic structure is still lacking. 

To generate the halo structure in heavier systems, from the viewpoint of the mean-field picture, if the valence neutron moves under a deformed potential, it would be more favorable for the occupation of $s$- or $p$-wave orbitals, which in turn facilitates the formation of deformed halos \cite{MISU1997,Zhou10, Hamamoto2012}. This implies that the condition for halo formation is expected to be less restrictive in deformed nuclei. If the projection $\Omega $ of the total angular momentum of the weakly bound valence levels is $\Omega=1/2$ or $3/2$, which might correspond to the occupation of $l=0$ or 1 orbitals, a halo can consequently be formed. Along an isotopic chain, the one-neutron separation energies of odd-$N$ nuclei typically vanish before even-$N$ nuclei reach the neutron drip line, mainly due to the blocking effect. Therefore, it will be experimentally easier to find neutron-halo nuclei among odd-$N$ isotopes. Moreover, the scheme of low-lying levels in even-$Z$ odd-$N$ nuclei would directly reflect the underlying shell structure. Thus, in the present work, we limit ourselves to the study of deformed one-neutron halos in nuclei with even-$Z$ and odd-$N$ in the mass region with $40<A<90$.  
 
The halo structure can be identified experimentally, for example, by measuring either a larger radius of the matter distribution or detecting a significant probability of weakly bound neutrons far from the well-bound core nuclei. The typical experiments for the former quantity are to measure reaction cross-sections, while Coulomb breakup reactions are employed for the latter, in which the presence of a halo can induce a large concentration of electric dipole ($E1$) strength in the low-excitation energy region, referred to as soft dipole excitation \cite{NAKAMURA1997,Nakamura1999,Nakamura2006}. In medium-heavy nuclei, the Coulomb breakup reaction will be more efficient since the matter radius can be affected by other factors, such as large deformation. Moreover, the ratio of the number of halo particles to that of nucleons in a well-bound core nucleus decreases in heavier systems, making the impact of the halo orbit on the total matter radius less pronounced than in lighter halo nuclei. In Refs.~\cite{Nakamura2009,Nakamura2014}, the Coulomb breakup cross sections for $^{31}$Ne were measured by Nakamura \textit{et al.}, indicating a soft dipole excitation in this nucleus. It has been shown in Ref.~\cite{Nakamura2014} that the deformation plays an important role in the formation of the $p$-wave halo in $^{31}$Ne due to the mixing of $pf$ shells.

Theoretically, the structural information of deformed halo nuclei has been widely studied within the deformed Woods-Saxon model \cite{Hamamoto2005,Hamamoto2010,Hamamoto2012,Hamamoto2017}, the particle-rotor model \cite{Urata2011,Urata2012,Urata2013}, the shell model \cite{Otsuka93,Kuo97}, the cluster model \cite{DESCOUVEMONT1999}, the deformed relativistic Hartree-Bogoliubov theory in continuum (DRHBc) \cite{Zhou10,li12,Li:2012xaa,SUN18,zhang19,SUN20,SUN21,SUN20212072,Sun:2021nfb,17B21,Zhong2022,ZhangKY23,ZHANG2023138112,plbzhang2024,Pan:2024qkc,Zhang2024PRC,Wang2024EPJA,Papakonstantinou2025PRC,Zhang2025PLB,Pan2026PLB,Zhang2026Particles}, the non-relativistic Hartree-Fock-Bogoliubov approach \cite{pei131,pei132,Nakada18,Kasyua2021,peisy21}, the anti-symmetrized molecular dynamics model \cite{Takatsu2023}, as well as a very recent lattice effective field theory study \cite{Shen:2024qzi}. Among these, the DRHBc framework is particularly noteworthy as it self-consistently accounts for deformation, continuum effects, pairing correlations, and their mutual coupling, all of which are essential for properly describing the halo phenomenon (see reviews \cite{Meng_2015,Sun2024_NPR,Zhang2024NPR,Zhang2025AAPPS} and references therein). 

In this paper, we employ the DRHBc theory to investigate  the effect of intrinsic deformation on the soft dipole excitation of odd-$N$ nuclei in the $40<A<90$ mass region. 
In particular, we select $^{43}$Si, $^{69}$Ti, and $^{75}$Cr as promising candidates for deformed halos. According to the DRHBc mass table~\cite{DRHBc-mass2024}, these nuclei meet the essential criteria for halo formation: They have small one-neutron separation energies ($S_n < 1$~MeV) and valence neutrons occupying $\Omega^\pi = 1/2^{+/-}$ states, which favor the dominance of low angular momentum ($l=0$ or 1) components. Furthermore, these nuclei are predicted to exhibit diverse ground-state shapes, with $^{43}$Si being oblate, while $^{69}$Ti and $^{75}$Cr are prolate.

This paper is organized as follows: Section \ref{model} is devoted to the theoretical framework of the DRHBc model and the soft dipole response calculation.
The results and discussions are presented in Section \ref{RD}. The summary is given in Section \ref{SUM}.

\section{Theoretical Framework}\label{model}
\subsection{The DRHBc theory}

The details of the DRHBc theory can be found in Refs.~\cite{Zhou10,li12}. 
In this section, we only briefly outline its theoretical framework. 
In the DRHBc theory, by treating the mean fields and pairing correlations self-consistently, the motion of nucleons is described by the relativistic Hartree-Bogoliubov (RHB) equation, 
\begin{equation} \label{RHB}
	\begin{pmatrix} \hat{h}_D - \lambda_\tau & \hat{\Delta} \\ -\hat{\Delta}^* & -\hat{h}_D^* + \lambda_\tau \end{pmatrix} 
	\begin{pmatrix} U_k \\ V_k \end{pmatrix} = E_k \begin{pmatrix} U_k \\ V_k \end{pmatrix}, 
\end{equation}
where $\hat{h}_D$ is the Dirac Hamiltonian, $\lambda_\tau$ is the Fermi energy ($\tau=n, p$ for neutrons or protons), $\hat{\Delta}$ is the pairing potential, and $E_k$ and $(U_k, V_k)^T$ are the quasiparticle energies and wave functions, respectively. 
After self-consistently solving the RHB equation \eqref{RHB}, the nucleon densities are obtained and the physical observables, such as binding energy, root-mean-square (rms) radii, and quadrupole deformation, can be calculated \cite{Zhang2020PRC,Pan2022PRC}. 
For an odd-$A$ nucleus, the blocking effect of the unpaired nucleon \cite{Ring1980NMBP} is included using the equal filling approximation \cite{Li:2012xaa,Pan2022PRC}. 

By diagonalizing the density matrix, one can obtain the occupation probabilities $v^2$ and the expansion coefficients of the canonical wave functions $\psi_{i(\Omega^\pi)}(\bm{r}s)$ in the spherical spinors of the Dirac Woods-Saxon (DWS) basis \cite{zhou03}: 
\begin{equation}\label{DRHBcwf}
    \psi_{i(\Omega^\pi)}(\bm{r}s) = \sum_{n\kappa} c_{n\kappa}^{i(\Omega^\pi)} \varphi_{n\kappa \Omega}(\bm{r}s). 
\end{equation}
Here, $\Omega$ is the third component of the total angular momentum $j$, $\pi$ is the parity, $n$ is the radial quantum number, and $\kappa=(-1)^{j+l+1/2}(j+1/2)$ is the relativistic quantum number.
The DWS basis is obtained by solving a Dirac equation with spherical Woods-Saxon potentials
\cite{zhou03}, and reads
\begin{equation}
\varphi_{n\kappa \Omega}(\bm{r} s)=\frac{1}{r}\left( \begin{matrix}
\mathrm{i} G_{n\kappa}(r) Y_{j\Omega}^l \\
-F_{n\kappa}(r) Y_{j\Omega}^{\tilde{l}}
\end{matrix}\right),
\end{equation}
where $G_{n\kappa}(r)/r$ and $F_{n\kappa}(r)/r$ are the radial wave functions for large and small components, $Y_{j\Omega}^{l}=\left[Y_{\ell} \otimes \chi_{1 / 2}\right]_{j \Omega}$ denotes the spinor spherical harmonics, and $\tilde{l}=l+(-1)^{j+l-1/2}$. 
 
In our calculations, we used the same numerical methods as those in the standard DRHBc mass table calculations \cite{Zhang2020PRC,Pan2022PRC}, with the exception of the box size. Since the DRHBc theory solves the RHB equation in a finite box, bulk properties of finite nuclei, such as binding energy and rms radii, are typically not influenced by the box size $R_\mathrm{Box}$ in practical calculations \cite{li12}. However, we note that near the box boundary, the radial wave functions of valence orbitals damp to zero quickly, 
especially in the case of wave functions with a halo nature.
This unphysical effect is corrected by  
the asymptotic behavior of the wave function.
We match the radial wave functions calculated with the box size $R_\mathrm{Box}=30$ fm to the asymptotic form at a designated radius $R_\mathrm{match}$, which is typically taken as $2\sim3$ times of the rms radius of the corresponding orbital. 
The asymptotic behavior for a bound state is given by \cite{Hamamoto2019024319}
\begin{equation}\label{APR}
    R(r) \propto r h_l^{(1)}(i\alpha r), \quad r \to \infty,
\end{equation} 
where $l$ is the orbital angular momentum quantum number, $\alpha = \sqrt{2\mu|\epsilon|}/\hbar$ with nucleon mass $\mu$ and single-nucleon energy $\epsilon$, and $h_l^{(1)}$ is the spherical Hankel function of the first kind. 

\subsection{{Soft dipole excitation of deformed one-neutron halo}} \label{SDR}

The single-neutron wave function in the continuum can be expressed within a plane wave approximation (PWA) for a  real energy variable $\varepsilon >0$ as 
\begin{equation}  \label{cwf}
\begin{aligned}
\left|\Phi_{\varepsilon}^{(c)}: \ell j m\right\rangle= & \frac{1}{r} R_{\ell j}^{(c)}(\varepsilon, r)\left[Y_{\ell} \otimes \chi_{1 / 2}\right]_{j m} \\
= & \sqrt{\frac{2 \mu}{\hbar^2 \pi k}}\left[\cos \left(\delta_{\ell j}\right) k j_{\ell}(k r)-\sin \left(\delta_{\ell j}\right) k n_{\ell}(k r)\right] \\
& \times\left[Y_{\ell} \otimes \chi_{1 / 2}\right]_{j m},
\end{aligned}
\end{equation}
where $j_{\ell}(k r)$ and $n_{\ell}(k r)$ are the spherical Bessel functions of the first kind and the second  kind, respectively, and $\delta_{\ell j}$ is the  phase shift. $k$ is defined by $k^2 = 2 \mu \varepsilon / \hbar^2$, with the reduced mass $\mu = (A-1)M/A$ for a system with mass number $A$.   
The radial wave function satisfies the normalization condition
\begin{equation}\label{radialcon}
\int_0^{\infty} d r R_{\ell j}^{(c)*}(\varepsilon, r) R_{\ell j}^{(c)}\left(\varepsilon^{\prime}, r\right)=\delta\left(\varepsilon-\varepsilon^{\prime}\right) .
\end{equation}

The electric dipole transition strength $I_i^\pi, K_i^\pi\rightarrow I_f^\pi, K_f^\pi$ is expressed in the laboratory frame as \cite{BM2}
\begin{equation}  \label{B-EL}
\begin{aligned}
& T\left(E 1 (\varepsilon); I_i^\pi, K_i^\pi \rightarrow I_f^\pi, K_f^\pi\right) \\
& =\left(e_{\mathrm{eff}}^n(E 1)\right)^2\left\{\left\langle I_iK_i1K_f-K_i|I_fK_f\right\rangle \right. \\
& \quad \times\left \langle K_f^\pi (\varepsilon)| r Y_{1\nu=K_f-K_i} |K_i^\pi\right \rangle \\
& \quad+(-1)^{I_i+K_i}  \left\langle I_i-K_i1K_f+K_i|I_fK_f\right\rangle   \\
& \left.\quad \times\left \langle K_f^\pi (\varepsilon) \mid r Y_{1\nu=K_i+K_f} \mid\widetilde{K_i^\pi}\right\rangle\right\}^2.
\end{aligned}
\end{equation}
It should be noted that, since the radial wave functions in the continuum are energy-normalized according to Eq.~\eqref{radialcon}, the reduced transition probability $T(E 1(\varepsilon))$ in Eq.~\eqref{B-EL} carries the dimension of $e^2~\text{fm}^2~\text{MeV}^{-1}$. Consequently, it represents the transition strength per unit energy, rather than a discrete transition probability.

For a single-nucleon halo nucleus, assuming the core is frozen during the process, the initial state can be expressed by the canonical wave function $\psi_{i(\Omega^\pi)}$ in Eq.~\eqref{DRHBcwf} with $K_i^\pi=\Omega^\pi$. Note that only the large component of the radial wave function is adopted in the practical calculations. It accounts for about 99\% of the total normalization weight, thereby justifying its use in the analysis of halo properties.
The final state is given by the continuum wave function [cf. Eq.~\eqref{cwf}] with $I_f=j$ and $K_f=m$.   
The effective electric charge of neutrons in the center-of-mass frame is given by
\begin{equation} \label{eff}
e_{\mathrm{eff}}^n(E 1)=\frac{N_{c}Z_{v}-Z_{c}N_v}{A},
\end{equation}
where $N_c (N_v)$ and $Z_c(Z_v)$ are the neutron and proton numbers of the core (valence) particles, respectively. In stable nuclei, the core polarization effect induced by coupling to giant dipole resonances is important for a quantitative estimate of $E1$ transition strength \cite{BM2}.   However, for loosely-bound neutrons, this effect is expected to be weak due to the decoupling from the core excitation; thus, the polarization charge is neglected here. 

The dipole transition strength at the  excitation energy $\omega$ in the continuum is 
\begin{equation}\label{Bst}
\begin{aligned}
\frac{\mathrm{d} B(E1)}{\mathrm{d} \omega} =&\int \mathrm{d} \varepsilon \delta\left(\omega-\left(\varepsilon+S\right)\right) 
\\
&\times T\left(E 1 (\varepsilon); I_i^\pi, K_i^\pi \rightarrow I_f^\pi, K_f^\pi\right),
\end{aligned}
\end{equation}
where $S>0$ is the separation energy of the last occupied neutron and $\varepsilon$ is the energy of continuum state in Eq. \eqref{cwf}.

\section{Results and discussions}\label{RD}
\subsection{Deformed single-neutron halo candidates in the $40 < A < 90$ region}

\begin{table*}[!htbp]
\centering
\small
\setlength{\tabcolsep}{12pt} 
\renewcommand{\arraystretch}{1.2}
\caption{Ground state properties, including quadrupole deformation ($\beta_2$), binding energy (BE), and one-neutron separation energy ($S_{n}$), calculated using the DRHBc theory with the PC-PK1 functional~\cite{DRHBc-mass2024} and the FRDM~\cite{Moller2016}. The notation $\Omega^\pi$ represents the last occupied neutron orbit in DRHBc calculations. The last two columns summarize the deformed halo status from the DRHBc theory and the deformed Woods-Saxon model by Hamamoto~\cite{Hamamoto2017}, respectively: circles ($\bigcirc$) denote deformed halo candidates, triangles ($\triangle$) represent candidates satisfying a relaxed condition ($1 < S_n < 2$~MeV), crosses ($\times$) indicate non-halo cases, and dashes (---) indicate that the nucleus was not studied. See the text for details.} \label{tab-odd-N}
\begin{tabular}{lcccccccccc}
\toprule
\toprule
 & & \multicolumn{3}{c}{FRDM} & \multicolumn{4}{c}{DRHBc} & \multicolumn{2}{c}{Deformed Halo}\\
\cmidrule(r){3-5} \cmidrule(lr){6-9} \cmidrule(l){10-11}
$A$ & $N$ & $\beta_2$ & $\rm BE$ & $S_n$ & $\beta_2$ & $\rm BE$ & $S_n$ & $\Omega^\pi$ & DRHBc & Ref.~\cite{Hamamoto2017}\\
 & & & (MeV) & (MeV) & & (MeV) & (MeV) & & & \\
\midrule
\multicolumn{11}{l}{$Z=14$ (Si)} \\
41 & 27 & $-0.302$ & 309.42 & 2.37 & $-0.339$ & 310.57 & 2.76 & $1/2^-$ & $\times$ & --- \\
43 & 29 & $-0.275$ & 312.63 & $-1.37$ & $-0.357$ & 316.08 & 0.95 & $1/2^-$ & $\bigcirc$ & --- \\
45 & 31 & --- & --- & --- & $-0.298$ & 318.89 & 0.41 & $1/2^-$ & $\bigcirc$ & --- \\
47 & 33 & --- & --- & --- & 0.000 & 321.54 & 0.34 & $1/2^-$ & $\times$ & --- \\
\midrule
\multicolumn{11}{l}{$Z=16$ (S)} \\
47 & 31 & $-0.238$ & 358.07 & 0.37 & $-0.226$ & 359.54 & 1.90 & $1/2^-$ & $\triangle$ & --- \\
49 & 33 & $-0.280$ & 361.23 & 0.04 & 0.000 & 364.83 & 1.82 & $1/2^-$ & $\times$ & --- \\
51 & 35 & $-0.280$ & 362.34 & $-1.09$ & $-0.093$ & 368.60 & 0.91 & $5/2^-$ & $\times$ & --- \\
53 & 37 & --- & --- & --- & $-0.095$ & 371.21 & 0.32 & $3/2^-$ & $\times$ & --- \\
55 & 39 & --- & --- & --- & 0.053 & 373.73 & 0.26 & $5/2^-$ & $\times$ & --- \\
\midrule
\multicolumn{11}{l}{$Z=18$ (Ar)} \\
51 & 33 & $-0.281$ & 402.17 & 1.45 & $-0.239$ & 405.16 & 2.81 & $5/2^-$ & $\times$ & --- \\
53 & 35 & $-0.280$ & 405.99 & $-0.07$ & $-0.209$ & 411.12 & 2.12 & $1/2^-$ & $\times$ & $\bigcirc$ \\
55 & 37 & $-0.213$ & 408.99 & 0.00 & $-0.153$ & 416.17 & 1.63 & $3/2^-$ & $\triangle$ & --- \\
57 & 39 & $-0.213$ & 411.02 & $-0.66$ & $-0.053$ & 420.33 & 1.40 & $1/2^-$ & $\times$ & --- \\
\midrule
\multicolumn{11}{l}{$Z=22$ (Ti)} \\
65 & 43 & 0.021 & 503.35 & 0.81 & $-0.028$ & 508.70 & 0.44 & $9/2^+$ & $\times$ & --- \\
67 & 45 & 0.129 & 507.46 & 0.45 & 0.214 & 512.78 & 0.74 & $5/2^+$ & $\times$ & --- \\
69 & 47 & 0.128 & 510.79 & 0.29 & 0.231 & 516.22 & 0.62 & $1/2^+$ & $\bigcirc$ & $\times$ \\
71 & 49 & 0.043 & 513.74 & 0.01 & 0.199 & 518.68 & 0.15 & $1/2^+$ & $\bigcirc$ & --- \\
\midrule
\multicolumn{11}{l}{$Z=24$ (Cr)} \\
71 & 47 & 0.172 & 553.79 & 1.12 & 0.291 & 558.06 & 1.99 & $1/2^+$ & $\triangle$ & $\bigcirc$ \\
73 & 49 & 0.053 & 559.00 & 1.35 & 0.314 & 562.27 & 1.26 & $7/2^+$ & $\times$ & $\bigcirc$ \\
75 & 51 & 0.032 & 561.57 & $-0.49$ & 0.311 & 565.07 & 0.46 & $1/2^+$ & $\bigcirc$ & $\bigcirc$ \\
77 & 53 & --- & --- & --- & 0.268 & 567.50 & 0.18 & $1/2^+$ & $\bigcirc$ & --- \\
\midrule
\multicolumn{11}{l}{$Z=26$ (Fe)} \\
75 & 49 & 0.053 & 600.33 & 2.47 & 0.228 & 600.95 & 2.29 & $1/2^+$ & $\times$ & $\times$ \\
77 & 51 & 0.042 & 604.76 & 0.24 & 0.282 & 605.63 & 1.33 & $1/2^+$ & $\triangle$ & $\bigcirc$ \\
79 & 53 & 0.011 & 606.84 & $-0.42$ & 0.255 & 609.62 & 1.00 & $1/2^+$ & $\bigcirc$ & --- \\
81 & 55 & 0.172 & 607.34 & $-1.26$ & 0.212 & 612.74 & 0.53 & $1/2^+$ & $\bigcirc$ & --- \\
83 & 57 & --- & --- & --- & 0.189 & 615.49 & 0.25 & $1/2^+$ & $\bigcirc$ & --- \\
85 & 59 & --- & --- & --- & 0.151 & 617.96 & 0.24 & $1/2^+$ & $\bigcirc$ & --- \\
87 & 61 & --- & --- & --- & $-0.072$ & 620.02 & 0.12 & $1/2^+$ & $\times$ & --- \\
\bottomrule
\bottomrule
\end{tabular}
\end{table*}
 
The ground state properties calculated by using the DRHBc theory \cite{DRHBc-mass2024}
with PC-PK1 \cite{PhysRevC.82.054319} are listed in Table \ref{tab-odd-N}, alongside the results from the finite-range droplet model (FRDM) in Ref. \cite{Moller2016}, for the isotopes of Si, S, Ar, Ti, Cr, and Fe, corresponding to $Z=14,$ 16, 18, 22, 24, and 26, respectively. 
Candidate deformed halo is indicated in the last two columns of Table \ref{tab-odd-N}. The FRDM results are listed for comparison of bulk properties; no halo assessment is made for FRDM as it does not provide orbital quantum number information.  We adopt three criteria for the selection of deformed halo nuclei:
\begin{itemize}
    \item The quadrupole deformation parameter satisfies $|\beta_2| \gtrsim 0.2$.
    \item The parity and the projection of the total angular momentum of the valence orbital, $\Omega^\pi$, are $1/2^+$, $1/2^-$, or $3/2^-$, which correspond to $s$- or $p$-wave halo components. 
    \item The $1n$ separation energy is less than or equal to $1$~MeV ($S_n \leq 1$~MeV). We also include halo candidates identified using a relaxed criterion ($1~\text{MeV} < S_n < 2~\text{MeV}$), denoted by the symbol "$\triangle$".  
\end{itemize}

We do not find deformed halo candidates 
in Ca ($Z=20$) and Ni ($Z=28$) isotopes due to the nature of semi-closed or closed proton shell in these isotopic chains.  
In general, the drip line in DRHBc calculations is more extended than FRDM. One of the important difference between FRDM and DRHBc is the coupling effect to the continuum states, which is  taken into account  self-consistently in DRHBc calculations. The continuum effect in DRHBc would make further extension of the neutron drip line compared with FRDM results. 
The DRHBc theory provides reasonable descriptions for the observed deformed halo nuclei $^{31}$Ne \cite{Zhong2022,Pan:2024qkc}
and $^{37}$Mg \cite{ZHANG2023138112}.
For heavier mass nuclei with $A>40$, the theory suggests possible oblate $p$-wave halos in $^{43}$Si and $^{45}$Si, and prolate $s$-wave halos in $^{69,71}$Ti, $^{75,77}$Cr, and $^{79,81,83,85}$Fe.

In Ref.~\cite{Li2024PRC}, Li \textit{et al.}~performed \textit{ab initio} valence-space in-medium similarity renormalization group (VS-IMSRG) calculations to predict neutron halo candidates in the Mg--S isotopic chains based on one neutron separation energies, spectroscopic factors, and two-nucleon amplitudes.  For the Si and S isotopic chains, they suggested $^{41,43,45,47}$Si and $^{47,49}$S as $1n$ halo candidates. Their predicted range is broader than that of the DRHBc theory mainly because they consider all halo candidates regardless of deformation, whereas we focus exclusively on deformed halos with $|\beta_2| \gtrsim 0.2$. In addition, the two approaches are based on different theoretical frameworks, leading to quantitative differences in the predicted separation energies. Nevertheless, their predictions of $^{43}$Si and $^{45}$Si are consistent with the DRHBc theory.
  
\begin{figure}[ht!]
\centering
\includegraphics[width= 8cm]{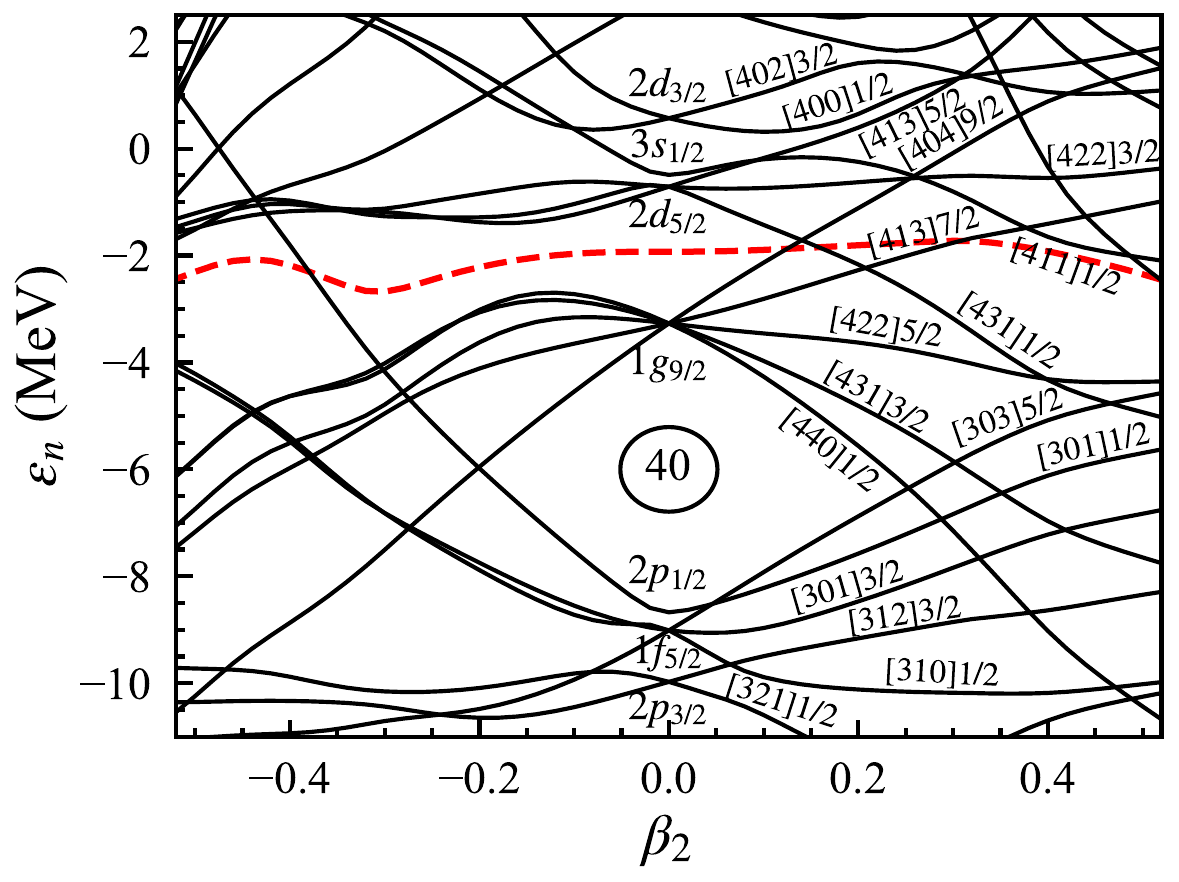}
\caption{(Color Online) Single neutron levels of $^{74}$Cr calculated using the DRHBc theory with PC-PK1 as a function of quadrupole deformation $\beta_2$. The Fermi surface is marked with a red dashed line. The asymptotic quantum numbers $[N n_z \Lambda]\Omega$ are denoted for the single-particle level, where $N$, $n_z$, $\Lambda$ and $\Omega$ represent the total oscillator quantum number, the oscillator quantum number in the $z$-direction, the orbital angular momentum projected on the symmetry axis, and the projection of total angular momentum on the symmetry axis, respectively.
\label{Cr75Nilsson}}
\end{figure}

In Ref.~\cite{Hamamoto2017},  Hamamoto studied possible deformed halo candidates for $A>50$ by using a deformed Woods-Saxon potential without pairing and continuum correlations. She identified $^{71}$Cr, $^{73}$Cr, $^{75}$Cr, and $^{77}$Fe as possible $s$-wave halo candidates. This was attributed to the narrowing of the $N = 50$ spherical energy gap near the neutron emission threshold, which induces significant deformation through the mixing of the $1g_{9/2}$, $3s_{1/2}$, and $2d_{5/2}$ neutron levels, as illustrated in Fig.~\ref{Cr75Nilsson}.   
This level structure around the $N=50$ Fermi level also appears in our calculations, where these 4 nuclei show large prolate deformation, while the DRHBc theory takes into account the pairing and continuum correlations in a self-consistent manner.  
The nucleus $^{53}$Ar is also predicted as a candidate of $s$- or $p$-wave halo with prolate deformation in Ref. \cite{Hamamoto2017}, while the DRHBc theory assigns a $\Omega^\pi=1/2^-$ state as the ground state of $^{53}$Ar with oblate deformation $\beta_2=-0.209$ and a separation energy $S_n=2.12$ MeV.  
The FRDM also predicts an oblate deformation for the ground state of $^{53}$Ar.  
 
\subsection{Density distribution and soft dipole excitations of deformed halo nuclei}
In the following, we select three nuclei $^{43}$Si, $^{69}$Ti, and $^{75}$Cr from Table \ref{tab-odd-N}, all of which are suggested to have deformed halo structures in our DRHBc calculations, while $^{43}$Si and $^{75}$Cr are slightly unbound in the  FRDM results.   The nucleus $^{75}$Cr is discussed as a deformed halo candidate in 
 the deformed Woods-Saxon model \cite{Hamamoto2017}.
According to the configuration assignments, $^{43}$Si is suggested to feature an oblate $p$-wave halo, while $^{69}$Ti and $^{75}$Cr exhibit prolate $s$-wave halos. In this part, we will analyze the halo structures of these three candidates one by one.

\subsubsection{$^{43}\mathrm{Si}$}
 
\begin{table}[htbp]
\centering
\setlength{\tabcolsep}{4pt}
\renewcommand{\arraystretch}{1.2}
\caption{Single-neutron orbitals near the Fermi surface of $^{43}$Si calculated with the DRHBc model. For each orbital $n$, the third component of the total angular momentum and parity $K^\pi$, single-particle energy $\varepsilon$, root-mean-square radius $R$, occupation probability $\nu^2$, and main components of the single-particle wave function are listed.}
\label{spl43si}
\begin{tabular}{cc ccc cccc}
\toprule
\toprule
\multirow{2.5}{*}{$n$} & \multirow{2.5}{*}{$\Omega^\pi$} & \multirow{2.5}{*}{$\varepsilon$ (MeV)} & \multirow{2.5}{*}{$R$ (fm)} & \multirow{2.5}{*}{$\nu^2$} & \multicolumn{4}{c}{Main Components} \\
\cmidrule(lr){6-9}
& & & & & $1f_{7/2}$ & $2p_{3/2}$ & $1f_{5/2}$ & $2p_{1/2}$ \\
\midrule
1 & $3/2^-$ & $-5.50$ & 4.77 & 1.0 & 32\% & 65\% & ---  & ---  \\
2 & $5/2^-$ & $-5.31$ & 4.38 & 1.0 & 78\% & ---  & 21\% & ---  \\
3 & $1/2^-$ & $-4.92$ & 4.82 & 1.0 & 30\% & 41\% & 1\%  & 24\% \\
4 & $1/2^-$ & $-1.22$ & 5.59 & 0.5 & 35\% & 1\%  & 3\%  & 58\% \\
\bottomrule
\bottomrule
\end{tabular}
\end{table}

The nucleus $^{43}$Si is a promising  candidate for a medium-heavy deformed neutron halo. Its ground state is assigned to the $K^\pi = 1/2^-$ state with a large oblate deformation ($\beta_2 = -0.357$) and Nilsson asymptotic quantum numbers $[Nn_z\Lambda]\Omega=[321]1/2$. The one neutron separation energy is predicted to be $S_n=0.95$ MeV. The single-neutron levels around the Fermi surface and their occupation probabilities are summarized in Table \ref{spl43si}. The pairing gap for this system is evaluated to be zero since  a large energy gap exists between the $n=4, K^\pi = 1/2^-$ and $n=3, K^\pi = 1/2^-$ configurations. The $1/2^-$ level labeled $n=4$ forms the neutron halo, while the other  bound states contribute to the neutron core. This halo orbital is dominated by the $2p_{1/2}$ component (58\%) and, due to the low centrifugal barrier, exhibits a significantly extended root-mean-square radius of $5.59$ fm. Recent research by Pan \textit{et al.}~\cite{Pan2026PLB} further highlighted a ``shape decoupling'' between the oblate core and the nearly spherical halo, leading to a unique ``rectangular box'' total density distribution. This structural prediction is further examined by their theoretical analysis of reaction observables, including predicted enhancements in reaction cross sections and narrow longitudinal momentum distributions~\cite{Pan2026PLB}. Building on these static and reaction insights, we will focus on the dynamical response of such a structure to investigate how this specific deformed halo configuration and its configuration mixing manifest in the $E1$ strength distribution.

\begin{figure}[ht!]
    \centering
    \includegraphics[width= 8 cm]{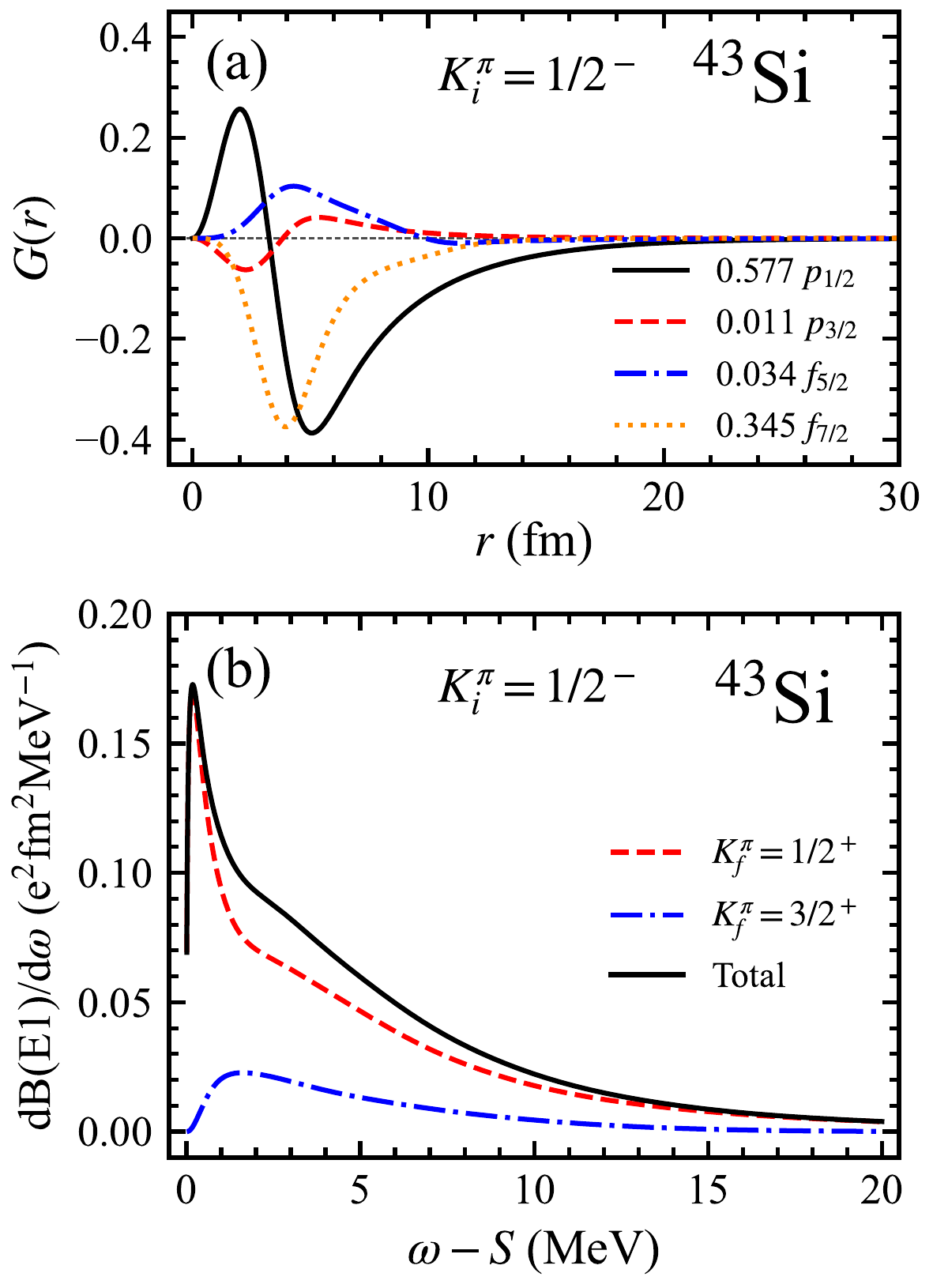}
    \caption{(Color Online) (a) The upper component of the radial Dirac wave functions of the halo neutron in $^{43}$Si calculated by the DRHBc theory, with the amplitudes of different components indicated. The asymptotic corrections [Eq. (\ref{APR})] are applied in the large-$r$ region. (b) Dipole strength distribution calculated  as function of the excitation energy (referred to the particle threshold). The initial state is given by the canonical wave functions of the last neutron [Eq. (\ref{DRHBcwf})] obtained from the DRHBc calculation, and the  final state is calculated in  the PWA.}
    \label{fig:si43dp1}
\end{figure}

\begin{figure}[ht!]
    \centering
    \includegraphics[width= 8 cm]{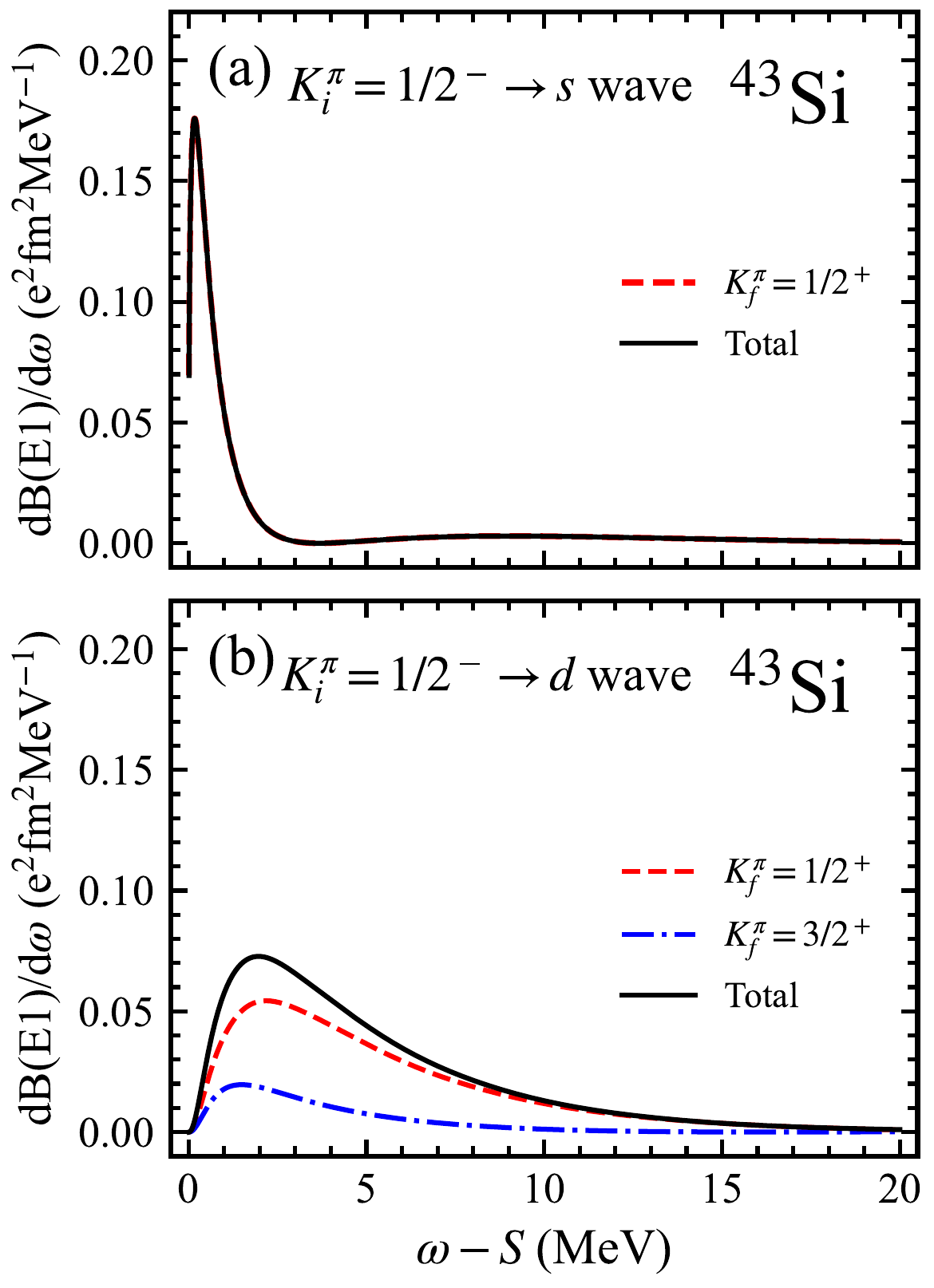}
    \caption{(Color Online) Dipole strength distribution in $^{43}$Si calculated under the same conditions as Fig. \ref{fig:si43dp1}(b). Panels (a) and (b) show the individual contributions from transitions to $s$-wave and $d$-wave final states, respectively. }
    \label{fig:si43dp2}
\end{figure}

The various components of the radial wave function for the last halo neutron configuration in $^{43}$Si are shown in Fig. \ref{fig:si43dp1}(a), with the amplitude of different $lj$ components indicated inside the panel. Note that only the upper components of the Dirac spinor are shown. For convenience, we plot $G(r)$ rather than the radial wave function $G(r)/r$ in the figures.
The $p$-wave halo component $p_{1/2}$ dominates the single-particle wave function of the halo neutron. The octupole components represented by the $f$-orbits  are also large as much as 40\%. The dipole excitation strength distribution, calculated using the DRHBc-derived halo neutron wave function with $K_i^\pi = 1/2^-$ as the initial state, is shown in Fig. \ref{fig:si43dp1}(b). The angular momentum of the initial state is assigned as $I_i^\pi=K_i^\pi=1/2^-$, while the final states are taken as $I_f^\pi=K_f^\pi=1/2^+$,  $I_f^\pi=3/2^+, K_f^\pi=1/2^+$,  and $I_f^\pi=3/2^+, K_f^\pi=3/2^+$.
The dipole response to $K_f^\pi=1/2^+$ shows a sharp peak structure just above  the threshold at $(\omega -S)\sim$0.3 MeV, which is a typical dipole response of the halo  $p$ state to  the continuum $s$-wave. The broad shoulder on the higher-energy side of the peak is dominated by  the  excitation  from the $p_{1/2}$-state to the continuum $d$-wave. 
This behavior is clearly illustrated in Fig. \ref{fig:si43dp2}, where the transition peak to the $s$-wave is located near the neutron emission threshold, and the peak for the $d$-wave transition appears at approximately 2 MeV. The peak position shown in Fig.~\ref{fig:si43dp1}(b) is shifted to higher energy by about 2 MeV due to the response associated with $f \rightarrow g$ and $f \rightarrow d$ transitions, which have essentially no peak structure due to the high centrifugal barrier, despite the large $f$-wave component listed in Table \ref{spl43si}. However, the shoulder strength in the upper part of the peak is certainly due to the octupole component of the halo wave function. 

\begin{table}[htb]
    \caption{Integrated $B(E1)$ strengths (in $e^2\text{fm}^2$) for $^{43}$Si from the threshold to $\omega-S = 5$~MeV, calculated for different initial configurations with $K_i^\pi = 1/2^-$. Values in parentheses represent the strengths without the time-reversed state contribution.}
    \label{tab:si43-5}
    \centering
    \setlength{\tabcolsep}{20pt}
    \renewcommand{\arraystretch}{1.1}
    \begin{tabular}{lcc}
        \toprule\toprule
        & \multicolumn{2}{c}{Initial configuration} \\ 
        \cmidrule(lr){2-3}
        $K_f^\pi$ & DRHBc & Pure $2p_{1/2}$ \\ 
        \midrule
        $1/2^+$ & 0.385 (0.128) & 0.439 (0.109) \\ 
        $3/2^+$ & 0.086         & 0.196         \\ 
        \midrule
        Total & \textbf{0.471 (0.214)} & \textbf{0.635 (0.305)} \\ 
        \bottomrule\bottomrule
    \end{tabular}
\end{table}
 
\begin{figure}[ht!]
    \centering
    \includegraphics[width= 8 cm]{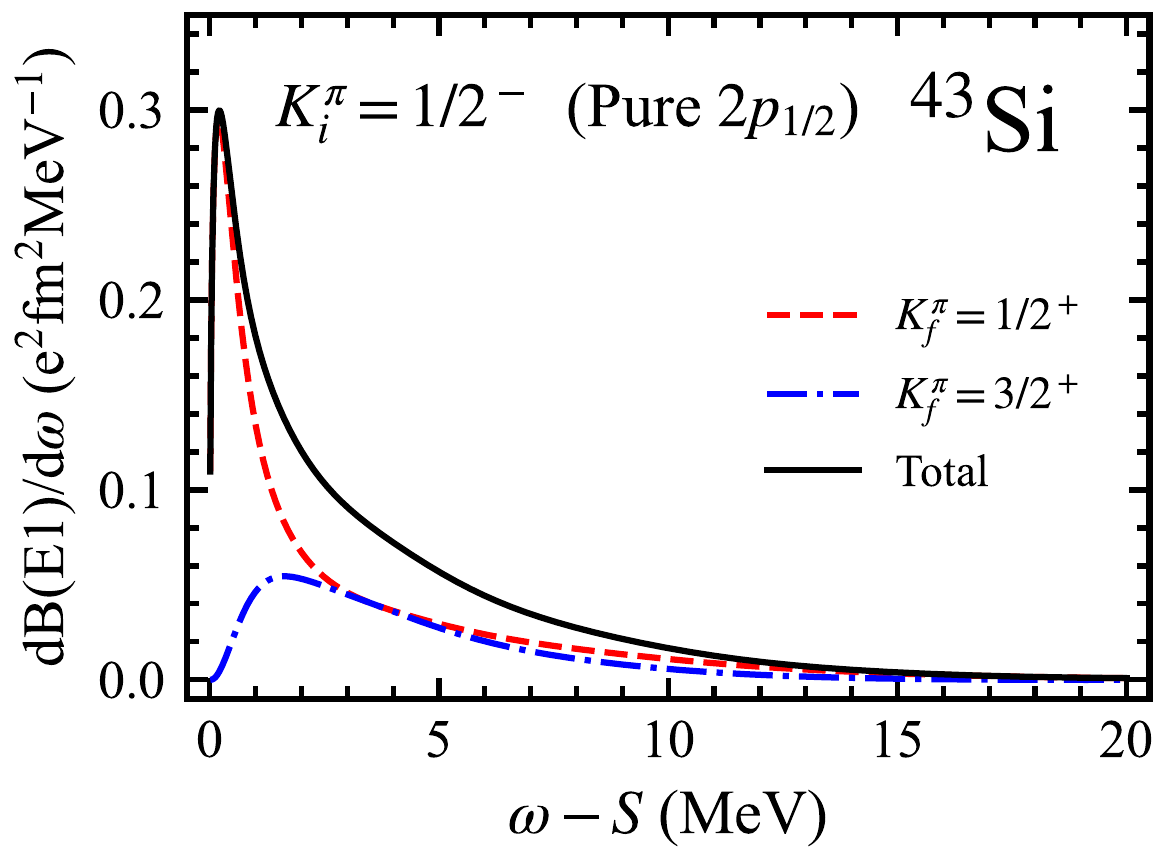}
    \caption{(Color Online) Dipole strength distribution in $^{43}$Si for a pure  $2p_{1/2}$  halo configurations. Initial state wave functions are calculated using the same spherical Woods-Saxon potential as in the DRHBc calculation.} 
    \label{fig:si43dp3}
\end{figure}

To further analyze the configuration dependence of the dipole excitation, we adopt a pure $2p_{1/2}$ wave function for the dipole
response calculation as shown in Fig. \ref{fig:si43dp3}. The $2p_{1/2}$ wave function is calculated using a spherical Woods-Saxon potential identical to that employed when generating the Dirac Woods-Saxon basis used in the DRHBc calculations.  The response of the $\Delta K=0$ transition has a sharp peak just above the threshold, similar to that obtained with the deformed configuration in Figs. \ref{fig:si43dp1}(b) and \ref{fig:si43dp2}(a), although  the peak height is approximately two times larger than that of the deformed configuration. This is simply because the transition from the halo $2p_{1/2}$ state to the $s_{1/2}$ continuum state is the origin of the peak, and the $2p_{1/2}$ component of the deformed configuration is 58\%, as shown in Table \ref{spl43si}.  
The response of the $\Delta K=1$ transition to the $K_f^\pi=3/2^+$ state exhibits almost the same energy dependence as the corresponding response in Fig. \ref{fig:si43dp1}(b), but the magnitude is again two times larger than that for the deformed case. 

The integrated $B(E1)$ values from the threshold up to an energy of 5 MeV are listed in Table \ref{tab:si43-5}. For the deformed configuration [321]1/2, the contributions from $K_f^\pi=1/2^+$ is approximately 4 times larger than that from $K_f^\pi=3/2^+$.  In an axial symmetric deformation,  an intuitive argument based on the degree of freedom of the allowed transitions in the axially-deformed nucleus predicts 2:1 ratio for the $\Delta K=0$ and $\Delta K=1$ transition strengths \cite{BM2}.  The large enhancement of the $\Delta K=0$ transition is due to the sharp threshold peak inherent in the halo wave functions, as clearly shown in Fig. \ref{fig:si43dp2}(a).   
Due to the significant mixing of higher-$l$ waves, the integrated $B(E1)$ strength of the deformed $K_i^\pi=1/2^-$ state calculated by DRHBc theory is quenched by approximately 25\% compared to that of the spherical $2p_{1/2}$ halo configuration. Nevertheless, the enhanced dipole strength  will  provide an opportunity to identify the deformation and the specific configuration of the  deformed halo in $^{43}$Si.

For dipole transitions, the second term in Eq. (\ref{B-EL}), coming from the time-reversed state $|\widetilde{K_i^\pi}\rangle$, contributes together with the $K_i=1/2$ state. Since the deformed halo configurations often have the quantum number $K_i^\pi=1/2^{\pm}$, this term should be examined carefully.  In the deformed halo nucleus $^{27}$Ne, this contribution was evaluated to be negligible for the dipole response of the [330]1/2 configuration because of a large cancellation between the major components in the  expansion of the Nilsson wave function in terms of the spherical basis \cite{Hamamoto2019}.  However, in the case of the deformed halo configuration [321]1/2 in $^{43}$Si, the second term makes a significant contribution, even larger than the first term in the integrated $B(E1)$ value shown in Table \ref{tab:si43-5}. In the DRHBc calculation, the time-reversed state gives a contribution approximately two times  larger than that from the $K_i^\pi=1/2^{-}$ state.

\begin{figure}[htb]
\centering
\includegraphics[width= 7.5 cm]{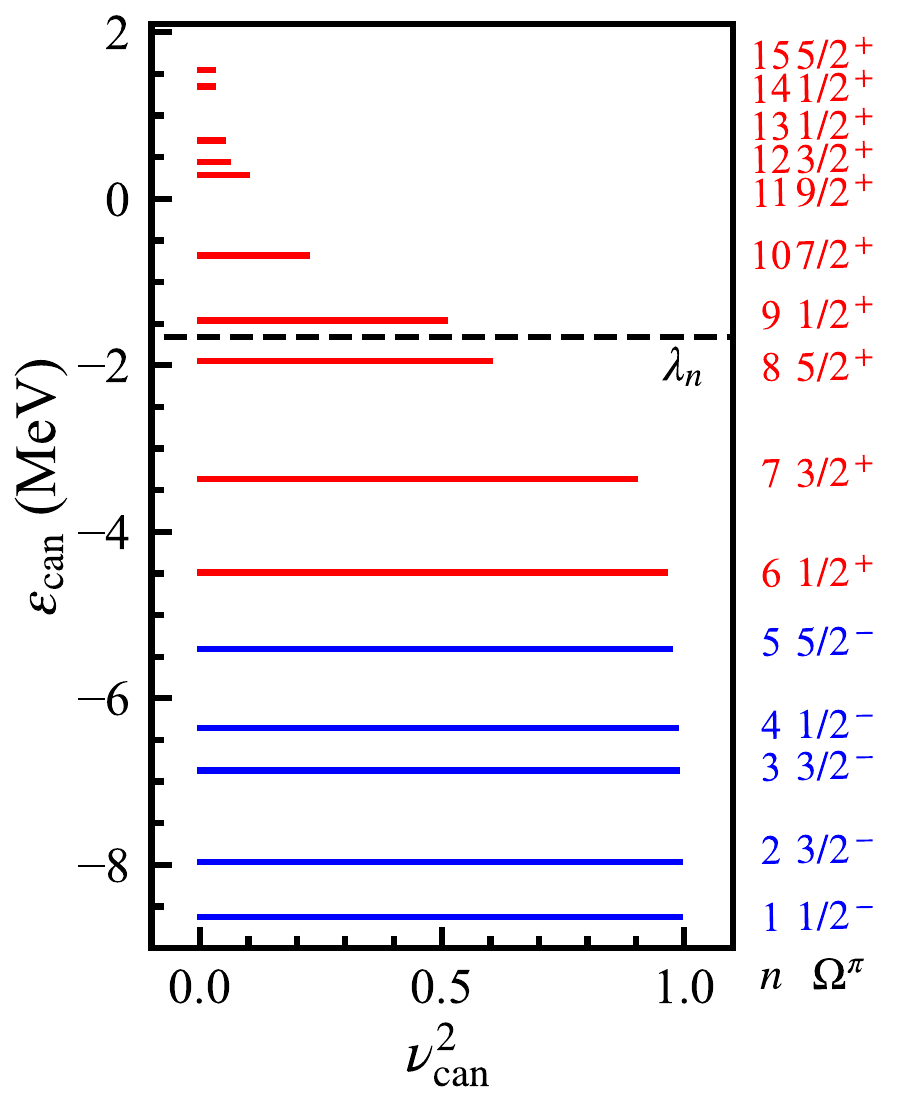} 
\caption{(Color Online) Single neutron levels versus occupation probabilities around the Fermi energy $\lambda_n$ in the canonical basis for $^{69}$Ti. The order $n$ and quantum numbers $\Omega^\pi$ are provided on the right side. The blue lines indicate the negative parity states, while the red color for positive parity.  \label{fig:ti69ev}} 
\end{figure}

\begin{table*}[htbp]
\centering
\renewcommand{\arraystretch}{1.2}
\caption{Same as Table \ref{spl43si}, but for $^{69}$Ti. The index $n$ corresponds to the labels used in Fig.~\ref{fig:ti69ev}.}
\label{spl69ti} 
\begin{tabular}{cc ccc cccccccccccccc}
\toprule
\toprule
\multirow{2.5}{*}{$n$} & \multirow{2.5}{*}{$\Omega^\pi$} & \multirow{2.5}{*}{$\varepsilon$ (MeV)} & \multirow{2.5}{*}{$R$ (fm)} & \multirow{2.5}{*}{$\nu^2$} & \multicolumn{14}{c}{Main Components} \\
\cmidrule(lr){6-19}
& & & & & $1g_{7/2}$ & $1g_{9/2}$ & $2d_{3/2}$ & $2d_{5/2}$ & $2g_{7/2}$ & $3s_{1/2}$ & $3d_{3/2}$ & $3d_{5/2}$ & $4s_{1/2}$ & $4d_{3/2}$ & $4d_{5/2}$ & $5s_{1/2}$ & $5d_{3/2}$ & $5d_{5/2}$ \\
\midrule
6  & $1/2^+$ & $-4.49$ & 5.32 & 0.96 & ---  & 65\%  & ---  & 19\% & ---  & 2\%  & 1\%  & 3\%  & 1\%  & ---  & 1\%  & 1\%  & ---  & 1\%  \\
7  & $3/2^+$ & $-3.36$ & 5.22 & 0.90 & 2\%  & 82\%  & ---  & 10\% & ---  & ---  & ---  & 1\%  & ---  & ---  & 1\%  & ---  & ---  & ---  \\
8  & $5/2^+$ & $-1.94$ & 5.14 & 0.60 & 1\%  & 93\%  & ---  & 3\%  & ---  & ---  & ---  & ---  & ---  & ---  & ---  & ---  & ---  & ---  \\
9  & $1/2^+$ & $-1.46$ & 6.76 & 0.51 & 1\%  & 22\%  & 4\%  & 14\% & ---  & 31\% & 11\% & 1\%  & 8\%  & 3\%  & ---  & 2\%  & 1\%  & ---  \\
10 & $7/2^+$ & $-0.68$ & 5.07 & 0.22 & 1\%  & 99\%  & ---  & ---  & ---  & ---  & ---  & ---  & ---  & ---  & ---  & ---  & ---  & ---  \\
11 & $9/2^+$ & $0.29$  & 5.03 & 0.10 & ---  & 100\% & ---  & ---  & ---  & ---  & ---  & ---  & ---  & ---  & ---  & ---  & ---  & ---  \\
12 & $3/2^+$ & $0.44$  & 5.74 & 0.06 & ---  & 15\%  & 1\%  & 62\% & ---  & ---  & 4\%  & 8\%  & ---  & 1\%  & 4\%  & ---  & 1\%  & 2\%  \\
13 & $1/2^+$ & $0.70$  & 5.83 & 0.05 & 16\% & 5\%   & 5\%  & 31\% & 1\%  & 1\%  & 20\% & 5\%  & ---  & 7\%  & 2\%  & ---  & 2\%  & 1\%  \\
14 & $1/2^+$ & $1.35$  & 6.60 & 0.03 & 26\% & 4\%   & ---  & 12\% & 2\%  & 31\% & 3\%  & 3\%  & 6\%  & 2\%  & 2\%  & 2\%  & 1\%  & 1\%  \\
15 & $5/2^+$ & $1.55$  & 5.65 & 0.03 & 1\%  & 4\%   & ---  & 73\% & ---  & ---  & ---  & 10\% & ---  & ---  & 5\%  & ---  & ---  & 3\%  \\
\bottomrule
\bottomrule
\end{tabular}
\end{table*}

\begin{figure*}[htb]
    \includegraphics[width= 14 cm]{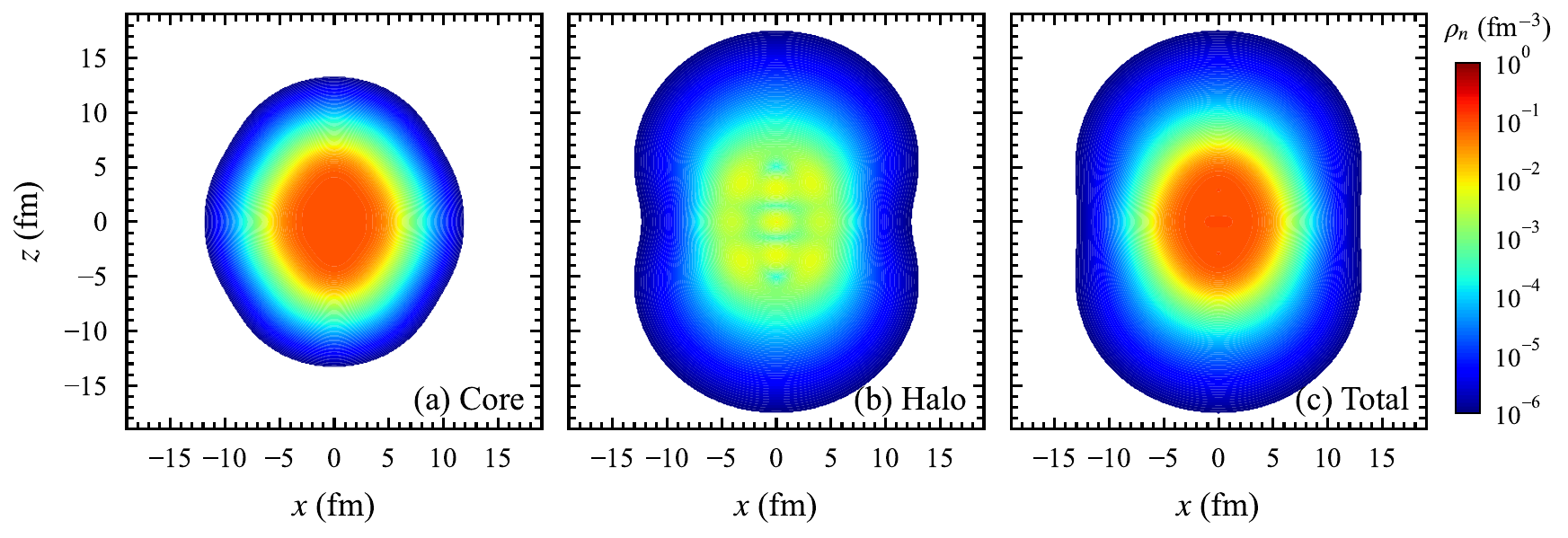}
    \caption{(Color Online) Two-dimensional neutron density distributions for $^{69}$Ti in the $xz$-plane from the DRHBc calculations with the PC-PK1. Panels (a), (b), and (c) correspond to the density distributions of the core, the halo, and the total neutron density, respectively. The $z$-axis is taken as the symmetry axis in the intrinsic frame, and the color scale indicates the density in fm$^{-3}$.\label{ti69den}}
\end{figure*}

\subsubsection{$^{69}\mathrm{Ti}$}

Based on the DRHBc calculations, the ground state of the nucleus $^{69}$Ti is assigned to the $K^\pi=1/2^+$ state, characterized by a large prolate deformation with $\beta_2 = 0.231$ and Nilsson asymptotic quantum numbers $[Nn_z\Lambda]\Omega = [420]1/2$ with $S_n = 0.62$ MeV. 
The single-particle levels near the Fermi surface and their occupation probabilities for $^{69}$Ti are depicted in Fig. \ref{fig:ti69ev}.  
As indicated by the occupation probabilities in  Fig. \ref{fig:ti69ev} and Table \ref{spl69ti}, this nucleus exhibits superfluid features in its ground state. Specifically, the pairing gaps are evaluated as $\Delta_n = 1.15$ MeV for neutrons and $\Delta_p = 1.11$ MeV for protons. The valence orbit $\Omega^\pi = 1/2^+$ consists of 42\% of the  $s_{1/2}$ halo component, along with 34\% of $d$-wave and 23\% of $g$-wave configurations.

Due to the small separation energy, the neutrons are distributed into many orbitals near the Fermi level. The continuum $n=9$ and $n=14$ levels with $\Omega^\pi=1/2^+$ contribute to the neutron halo, while the $n=8$, $\Omega^\pi=5/2^+$ level and other bound states contribute to the neutron core. The 2D density distributions of the core and halo wave functions, along with the total density, are shown in Fig. \ref{ti69den}. In Fig. \ref{ti69den}(b), an extended neutron halo density distribution along the $z$-axis is clearly seen.
Due to this peculiar halo density distribution, the total density exhibits a prolate rectangular shape extended in the $z$-direction, as seen in Fig. \ref{ti69den}(c).

\begin{figure}[htb]
    \centering
    \includegraphics[width= 8 cm]{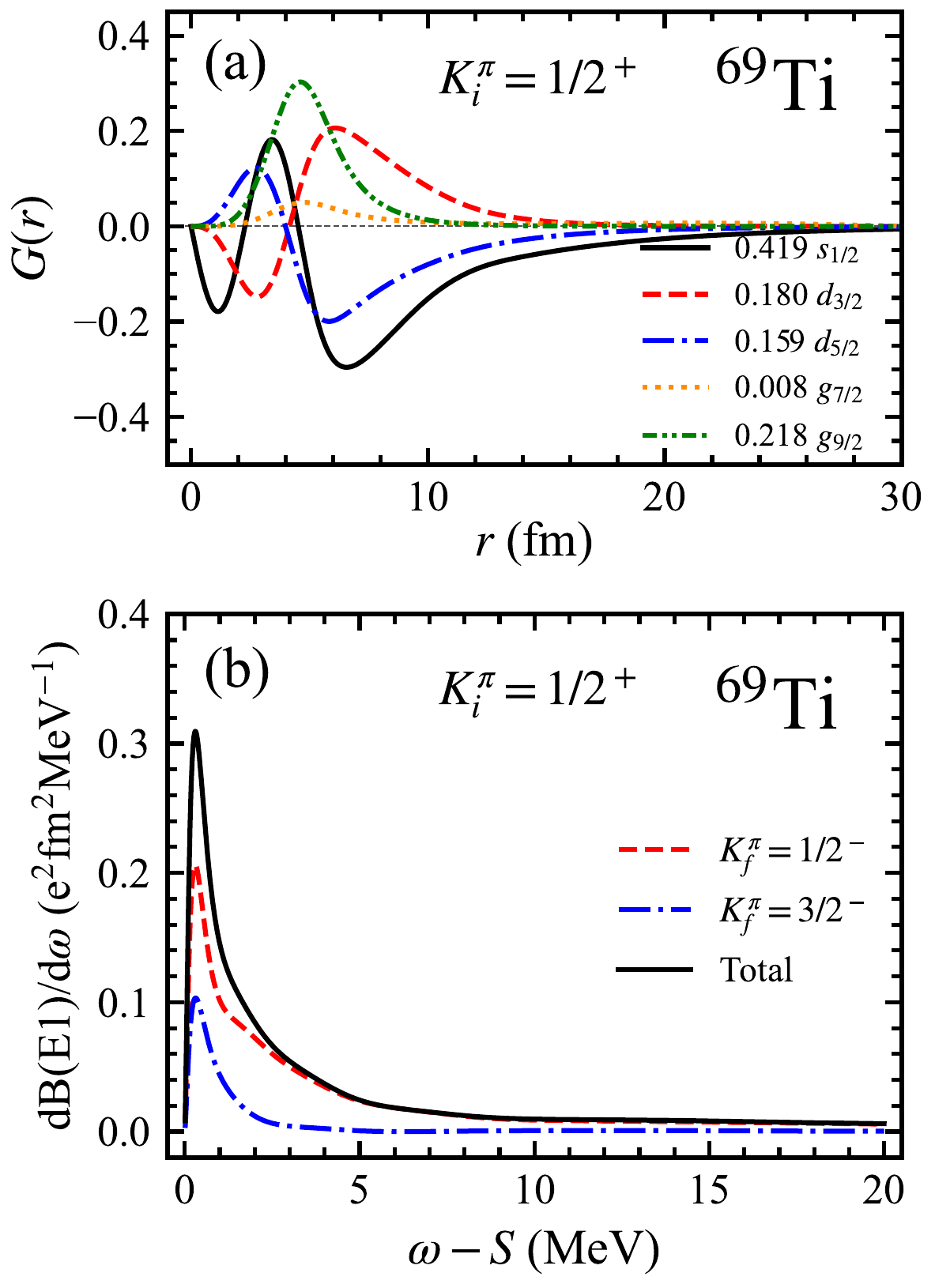}
    \caption{(Color Online) Same as Fig.~\ref{fig:si43dp1}, but for $^{69}$Ti. }
    \label{fig:ti69dp1}
\end{figure}

The radial wave functions and the electric dipole response for the prolate halo configuration in $^{69}$Ti are shown in Fig. \ref{fig:ti69dp1}. It is clearly seen in Fig. \ref{fig:ti69dp1}(a) that the $s_{1/2}$ halo component dominates the outer region,  $r > 10$ fm. 
In $^{69}$Ti, the transitions from the $K_i^\pi = 1/2^+$ ground state to both  $K_f^\pi = 1/2^-$ and  $K_f^\pi = 3/2^-$ channels create sharp peaks just above the threshold.  The integrated $B(E1)$ strengths are tabulated in Table \ref{tab:ti69-520}.  In the case of [420]1/2 configuration in $^{69}$Ti, the contribution from the time-reversed state dominates the transition to $K_f^\pi = 1/2^-$ and enhances the threshold peak.  As expected from the
 symmetry argument of deformation on the dipole transitions, the $\Delta K=0$ transition gives a larger contribution than the $\Delta K=1$ transition to the integrated $B(E1)$ value, as shown in Table \ref{tab:ti69-520}.

\begin{table}[ht!]
    \caption{Integrated $B(E1)$ strengths (in $e^2\text{fm}^2$) for $^{69}$Ti from the threshold to $\omega-S = 5$ and 20 MeV, calculated for the initial ground state with $K_i^\pi = 1/2^+$. Values in parentheses represent the strengths without the time-reversed state contribution. 
    }
    \label{tab:ti69-520}
    \centering
    \setlength{\tabcolsep}{20pt} 
    \renewcommand{\arraystretch}{1.2} 
    \begin{tabular}{lcc}
        \toprule
        \toprule
        & \multicolumn{2}{c}{Integration limit ($\omega-S$)} \\ \cmidrule(lr){2-3}
        $K_f^\pi$ & 5 MeV & 20 MeV \\
        \midrule
        $1/2^-$ & 0.361 (0.015) & 0.506 (0.116)\\
        $3/2^-$ & 0.108 & 0.116 \\
        \midrule
        Total   & \textbf{0.469  (0.123)} & \textbf{0.622 (0.233)} \\
        \bottomrule
        \bottomrule
    \end{tabular}
\end{table}

\subsubsection{$^{75}\mathrm{Cr}$}

\begin{table*}[htbp]
\centering
\setlength{\tabcolsep}{3pt} 
\renewcommand{\arraystretch}{1.2}
\caption{Same as Table \ref{spl43si}, but for $^{75}$Cr.}
\label{spl75cr}
\begin{tabular}{cc ccc ccccccccccccc}
\toprule
\toprule
\multirow{2.5}{*}{$n$} & \multirow{2.5}{*}{$\Omega^\pi$} & \multirow{2.5}{*}{$\varepsilon$ (MeV)} & \multirow{2.5}{*}{$R$ (fm)} & \multirow{2.5}{*}{$\nu^2$} & \multicolumn{13}{c}{Main Components} \\
\cmidrule(lr){6-18}
& & & & & $1g_{7/2}$ & $1g_{9/2}$ & $2d_{3/2}$ & $2d_{5/2}$ & $3d_{3/2}$ & $3d_{5/2}$ & $3s_{1/2}$ & $4d_{3/2}$ & $4d_{5/2}$ & $4s_{1/2}$ & $5d_{3/2}$ & $5d_{5/2}$ & $5s_{1/2}$ \\
\midrule
1  & $5/2^+$ & $-4.12$ & 5.15 & 0.96 & 3\%  & 89\% & ---  & 5\%  & ---  & ---  & ---  & ---  & ---  & ---  & ---  & ---  & ---  \\
2  & $1/2^+$ & $-3.39$ & 5.92 & 0.95 & 2\%  & 32\% & 18\% & 10\% & 5\%  & ---  & 22\% & 2\%  & ---  & 3\%  & 1\%  & ---  & 2\%  \\
3  & $7/2^+$ & $-1.93$ & 5.13 & 0.62 & 1\%  & 97\% & ---  & ---  & ---  & ---  & ---  & ---  & ---  & ---  & ---  & ---  & ---  \\
4  & $1/2^+$ & $-1.10$ & 6.49 & 0.50 & 23\% & 5\%  & 31\% & 26\% & 6\%  & ---  & 5\%  & 1\%  & ---  & ---  & ---  & ---  & ---  \\
5  & $3/2^+$ & $-0.73$ & 5.94 & 0.17 & 1\%  & 18\% & 10\% & 64\% & 3\%  & ---  & ---  & 1\%  & ---  & ---  & ---  & ---  & ---  \\
6  & $9/2^+$ & $-0.15$ & 5.06 & 0.13 & ---  & 99\% & ---  & ---  & ---  & ---  & ---  & ---  & ---  & ---  & ---  & ---  & ---  \\
7  & $1/2^+$ & $1.07$  & 5.72 & 0.03 & 44\% & 4\%  & 4\%  & 21\% & 1\%  & ---  & 17\% & ---  & ---  & 2\%  & ---  & ---  & 1\%  \\
8  & $5/2^+$ & $1.11$  & 5.72 & 0.03 & 1\%  & 6\%  & ---  & 87\% & ---  & 1\%  & ---  & ---  & 1\%  & ---  & ---  & 1\%  & ---  \\
9  & $3/2^+$ & $1.19$  & 5.54 & 0.03 & 71\% & ---  & 11\% & 10\% & 3\%  & ---  & ---  & 1\%  & ---  & ---  & 1\%  & ---  & ---  \\
10 & $1/2^+$ & $1.76$  & 6.07 & 0.02 & 25\% & 1\%  & 17\% & 10\% & 4\%  & ---  & 32\% & 1\%  & 1\%  & 3\%  & ---  & ---  & 2\%  \\
\bottomrule
\bottomrule
\end{tabular}
\end{table*}

\begin{figure*}[htb]
    \includegraphics[width= 14 cm]{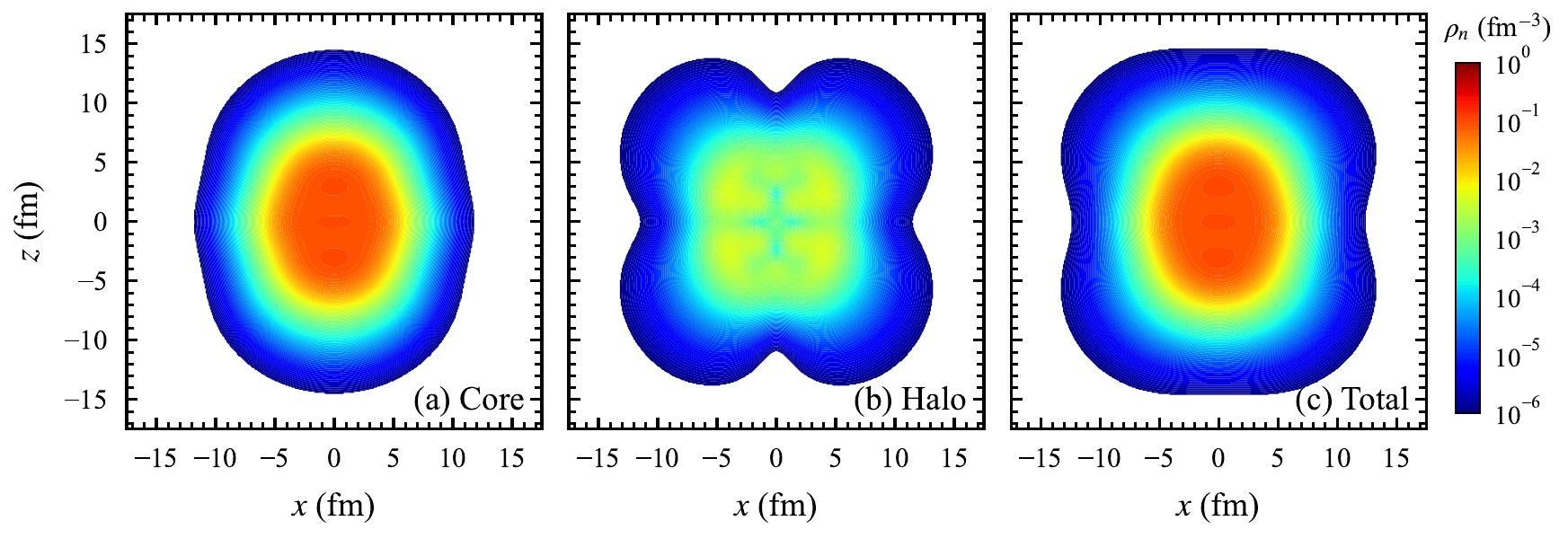}
    \caption{(Color Online) Same as Fig. ~\ref{ti69den}, but for $^{75}$Cr.\label{cr75den}}
\end{figure*}

\begin{figure}[htb]
    \centering
    \includegraphics[width= 8 cm]{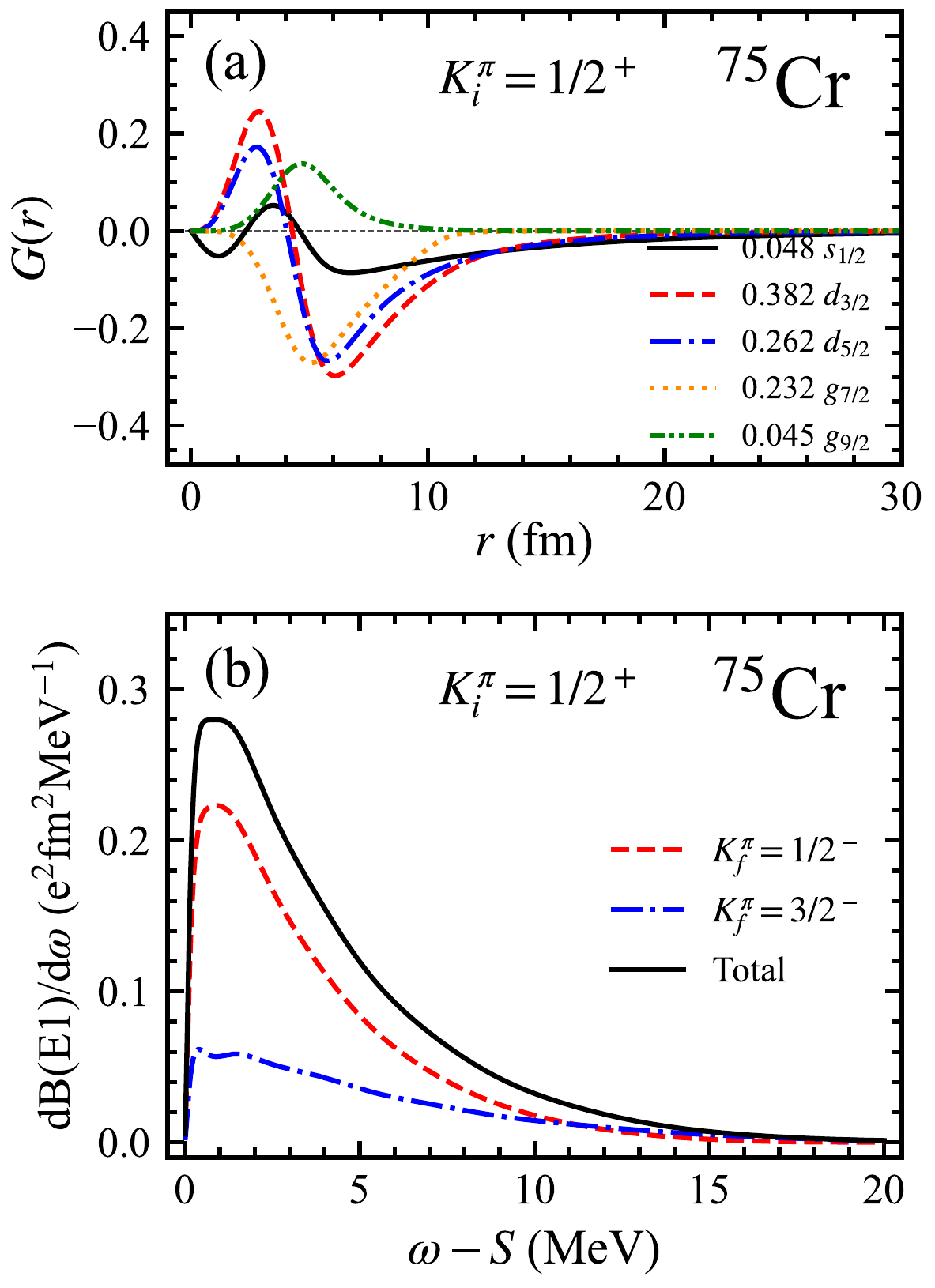}
    \caption{(Color Online) Same as Fig.~\ref{fig:si43dp1}, but for $^{75}$Cr. }
    \label{fig:Cr75dp1}
\end{figure}

The ground state of $^{75}\text{Cr}$ is assigned as the $K^\pi=1/2^+$ state, characterized by a prolate deformation of $\beta_2 = 0.311$ and corresponding to the Nilsson asymptotic quantum numbers $[Nn_z\Lambda]\Omega^\pi = [411]1/2^+$. Similar to $^{69}$Ti, $^{75}\text{Cr}$ also exhibits superfluid features in its ground state. The pairing gaps are evaluated as $\Delta_n = 0.97$ MeV for neutrons and $\Delta_p = 0.46$ MeV for protons. The one-neutron separation energy is $S_n = 0.46$ MeV. The configurations of the neutron levels around the Fermi surface are listed in Table \ref{spl75cr}, where the halo neutron orbital is labeled $n = 4$. This state is dominated by $2d$ and $1g$ configurations, with the $3s$ halo component accounting for only 5\%, suggesting a very weak halo configuration. Nevertheless, the halo density in Fig. \ref{cr75den}(b) shows an extension, which is a characteristic feature of an $s$-wave halo, and the radius for this halo level is notably large, as shown in Table \ref{spl75cr}. Furthermore, the substantial $1g$ component induces an intrinsic hexadecapole shape in the halo density. The combined core and halo density shown in Fig. \ref{cr75den}(c) exhibits a large extension in both the $x$- and $z$-directions.

The radial wave functions for the different components of the $[Nn_z\Lambda]\Omega = [411]1/2$ configuration in $^{75}\text{Cr}$ are shown in the Fig. \ref{fig:Cr75dp1}(a). As noted in Table \ref{spl75cr}, the $s_{1/2}$ halo component is as small as 5\%. However, it is evident that the halo component remains substantial in the tail region ($r > 10$ fm). The dipole response of $^{75}\text{Cr}$ is shown in Fig. \ref{fig:Cr75dp1}(b). Despite a small  halo component in the deformed a configuration $[411]1/2^+$, the dipole response exhibits a characteristic feature of  halo wave function; a strong peak just above the threshold having a large width $\Gamma_{\rm FWHM} \simeq 4$ MeV, which is much broader than those of $^{43}\text{Si}$ and $^{69}\text{Ti}$. This feature indicates that the dipole response is an extremely sensitive indicator of the halo component of the wave function. 

The integrated $B(E1)$ values are listed in Table \ref{tab:cr75-520} for energy limits of 5 and 20 MeV, respectively. In the case of $^{75}$Cr, the time-reversed state gives only a small contribution to the $B(E1)$ strength since the halo $s_{1/2}$ component is small. The response to the $K_f^\pi = 1/2^-$ final state  dominates the dipole response, as expected for $\Delta K=0$ transition, with a strength approximately  three times larger than that of the $K_f^\pi = 3/2^-$ channel in the low-energy region below 5 MeV. 
The ratio of the $\Delta K=0$ and $\Delta K=1$ transitions in the integrated $B(E1)$ value up to 20 MeV becomes  approximately 2.5.
This change certainly reflects the enhancement of the dipole strength near the threshold due to the halo component of the $[411]1/2$ level. 
  
\begin{table}[ht!]
    \caption{Same as Table \ref{tab:ti69-520}, but for $^{75}$Cr.  
    }
    \label{tab:cr75-520}
    \centering
    \setlength{\tabcolsep}{24pt} 
    \renewcommand{\arraystretch}{1.2} 
    \begin{tabular}{l cc}
        \toprule
        \toprule
        & \multicolumn{2}{c}{Integration limit ($\omega-S$)} \\ \cmidrule(lr){2-3}
        $K_f^\pi$ & 5 MeV & 20 MeV \\
        \midrule
        $1/2^-$ & 0.791 (0.742) & 1.053 (1.003) \\
        $3/2^-$ & 0.247 & 0.426 \\
        \midrule
        Total   & \textbf{1.038 (0.989)} & \textbf{1.479 (1.429)} \\
        \bottomrule
        \bottomrule
    \end{tabular}
\end{table}

\section{Summary}\label{SUM}
We studied possible deformed 1$n$ halo nuclei in the mass region $40<A<90$ using the DRHBc theory. Several deformed halo candidates are suggested, including $^{69,71}$Ti, $^{75,77}$Cr, and $^{79,81,83,85}$Fe with prolate deformations, as well as $^{43,45}$Si with oblate deformations.
The DRHBc results were compared with those from the FRDM and the deformed Woods-Saxon potential model. It was found that the DRHBc results show better agreement with the FRDM than with the deformed Woods-Saxon calculations. 
The halo density in these mass regions shows a variety of unique shapes induced by higher multipole components of the single-particle wave function, such as the octupole component ($f$-wave) in $^{43}$Si and the hexadecupole component ($g$-wave) in $^{75}$Cr.

The dipole response of three deformed halo candidates  $^{43}$Si, $^{69}$Ti, and  $^{75}$Cr are calculated using the wave functions obtained from the DRHBc theory. A large extension of halo wave function creates a unique nature of dipole response just above the neutron threshold.  
We found that the width of the sharp peak is highly dependent on the halo configuration of each nucleus; 
it is narrow in $^{43}$Si and ${^{69}}$Ti due to their large halo components, whereas it is broader in ${^{75}}$Cr because of large mixing of $d$- and $g$-wave components.  
The dipole response of $^{75}$Cr also shows a halo nature of sharp peak just  above the neutron threshold, although 
the  halo component of the deformed configuration [411]1/2 is small as 5\%.  The width of peak in $^{75}$Cr is the largest due to the significant mixing amplitudes of higher angular momentum components $l=2$ and $l=4$. Nevertheless, the peak structure in $^{75}$Cr clearly indicates that the dipole response is a highly sensitive observable for detecting deformed halo wave functions.  

It is generally shown that the dipole response of a deformed halo exhibits an inherent sharp peak structure just above the threshold,  and that
the width depends on the probability of halo component in the deformed wave function. These features of the dipole response may provide solid information about the deformed halo configuration, as well as the sign and magnitude of the nuclear deformation. Therefore, experimental data on the dipole response in the low-energy region of these deformed halo candidates are highly desirable from state-of-the-art  experimental facilities such as RIKEN (Wako), FRIB (MSU), and HIAF (Huizhou).

\begin{acknowledgments}
Helpful discussion with B. N. Lu, Y. M. Jiang, and members of the DRHBc Mass Table Collaboration are highly appreciated. X.-X. Sun acknowledges the Institute of Theoretical Physics for its hospitality and support during his visit, where part of this work was completed.
This work is supported by the National Key R\&D Program of China (Grant Nos. 2024YFE0109801 and 2023YFA1606503), the CAS Strategic Priority Research Program (Grant {Nos. XDB34010103 and XDB0920102}), the Chinese Academy of Sciences (CAS) President's International Fellowship Initiative (PIFI, Grant No. 2024PVA0003\_Y1) and the National Natural Science Foundation of China (Grant Nos. 12347139, {12447101}, 12375118, 12435008, 12205308, and W2412043). 
H.S. is supported by the JSPS Grant-in-Aid for Scientific Research (C) under Grant No. JP26K07079.
The results described in this paper are obtained on the High-performance Computing Cluster of ITP-CAS and the ScGrid of the Supercomputing Center, Computer Network Information Center of Chinese Academy of Sciences. 
\end{acknowledgments}
 
\bibliographystyle{apsrev4-2}
\bibliography{apssamp.bib} 

\end{document}